\documentclass{ws-procs975x65}

\begin{document}

\title{ACCRETING BLACK HOLES}

\author{MITCHELL C. BEGELMAN$^*$}

\address{JILA, University of Colorado and National Institute of Standards and Technology\\
440 UCB, Boulder, CO 80309-0440, USA\\
$^*$E-mail: mitch@jila.colorado.edu}

\begin{abstract}
I outline the theory of accretion onto black holes, and its application to observed phenomena such as X-ray binaries, active galactic nuclei, tidal disruption events, and gamma-ray bursts.  The dynamics as well as radiative signatures of black hole accretion depend on interactions between the relatively simple black-hole spacetime and complex radiation, plasma and magnetohydrodynamical processes in the surrounding gas.  I will show how transient accretion processes could provide clues to these interactions.  Larger global magnetohydrodynamic simulations as well as simulations incorporating plasma microphysics and full radiation hydrodynamics will be needed to unravel some of the current mysteries of black hole accretion.

\end{abstract}


\bodymatter

\section{The Spacetime Context}\label{aba:sec1}
Accretion of gas onto a black hole provides the most efficient means known for liberating energy in the nearby universe.  This is good for observational astronomers, since it allows regions close to the event horizon to be studied directly through emitted radiation.  But it is also crucial for understanding the influence of black holes on galaxy formation and evolution.  The huge amounts of energy released close to accreting black holes, particularly in the form of winds and jets, but also in the form of energetic radiation, can affect the thermal and dynamical states of matter out to large distances, making black holes important agents of change on cosmological scales.   

Astrophysical black holes are described by two parameters that effectively set the inner boundary conditions for accretion.  The mass, $M$, basically determines the characteristic length and time scales close to the horizon, whereas the Kerr spin parameter $a/M$ (where $a = J/M$ is the specific angular momentum of the hole, in geometric units $G = c = 1$), with $0 \leq a/M \leq 1$, determines the efficiency of energy release and, coupled to the magnetic and radiative properties of the infalling gas, the forms in which energy is liberated. 

Although the horizon (at $R_{\rm H}$) marks the point of invisibility and no return for matter being accreted by a black hole, the energy efficiency of accretion is determined somewhat farther out, near the {\it innermost stable circular orbit\/} (ISCO). A test particle orbiting inside this radius will be swallowed by the black hole without giving up any additional energy or angular momentum.  Because $R_{\rm ISCO}$ decreases from $6 M$ ($= 3 R_{\rm H}$) for a Schwarzschild black hole to $M$ ($= R_{\rm H}$) for a corotating orbit around an extreme ($a/M = 1$) Kerr hole, accretion is more efficient for a rotating hole than for a stationary one.  (Note, however, that $R_{\rm ISCO}$ increases with $a/M$ for counter-rotating orbits and approaches $9M$ for extreme Kerr, affording much lower efficiency.)  The specific angular momentum corresponding to the ISCO decreases from $2\sqrt{3}M$ to $2 M/\sqrt{3}$ as $a/M$ increases from 0 to 1, and the efficiency of energy release increases from about 6\% of $M c^2$ to about 42\%.  Hartle's undergraduate textbook on relativity\cite{hart03} provides a very readable discussion of these key features.  

\section{The Gaseous Environment}
It is important to keep in mind that elements of gas in an accretion flow do not behave exactly as test particles close to the ISCO, and in some cases their dynamics can be influenced strongly by pressure and magnetic forces.  Gas (or radiation) pressure forces, directed inward, can allow gas to remain in orbit slightly closer to the black hole than the ISCO, and with somewhat higher angular momentum.  Gas plunging into the black hole from such an orbit would have a lower binding energy and therefore a lower accretion efficiency.  In the limiting case where gas orbits a Schwarzschild black hole down to $4M$ (the {\it marginally bound orbit}), the binding energy of the accreted gas approaches zero and so does the accretion efficiency\cite{kozl78}. 

Likewise, net magnetic flux, accumulating in the innermost regions of an accretion flow, could hold back the gas, creating a {\it magnetically arrested disk}\cite{nara03}.  The angular momentum close to the black hole might then be lower than that of any stable test particle orbit, but infall could be regulated by interactions between the gas and the magnetic field, such as interchange instabilities and reconnection. 

In light of these considerations, we can identify at least four factors that must play an important role in governing black hole accretion flows.  The first three of these may be regarded as outer boundary conditions for the problem. 

\begin{itemlist}
\item \underbar{\it Angular momentum}. The specific angular momentum at the marginally bound orbit, somewhat larger than $\ell_{\rm ISCO}$ but still of order a few $GM/c$, represents the largest angular momentum per unit mass that can be accreted by a black hole.  Given that the radius of the gas reservoir supplying black hole accretion is usually at least a few hundred times $M$, and often much more, accretion is seldom possible without the loss of some angular momentum.  Angular momentum is thought to be transferred outward through the flow via the magnetorotational instability (MRI)\cite{balb98}, which works in the limit of sufficiently {\it weak\/} magnetic field.  Angular momentum can also be lost through winds or electromagnetic torques; the latter process depends on net magnetic flux, the third factor below.

\item \underbar{\it Radiative efficiency}.  Energy liberated during the accretion process is thought to be transferred outward by the same processes that wick away the excess angular momentum.  Some of this energy can go into driving coherent motions such as circulations or outflows, or can be removed by electromagnetic torques.\cite{blan82}.  But much (or most) of it is likely to be dissipated as heat or radiation.  If much of this energy is retained by the flow, the associated pressure can partially support the flow against gravity, reducing the relative importance of rotation.  When the local rotation rate is still a large fraction of the Keplerian value, the flow resembles a disk but the added pressure force can drive circulation or even mass loss\cite{blan99,bege12}. When pressure support dominates rotational support, or even becomes comparable to it, the flow can inflate into a nearly spherical configuration and interesting stability questions arise. 
     
\item \underbar{\it Magnetic flux}.  The MRI basically uses distortions of a poloidal magnetic field (i.e., parallel to the rotation axis) to extract free energy from differential rotation\cite{balb98}.  Therefore, it is not surprising that the vigor of angular momentum transport driven by MRI is sensitive to the presence of poloidal magnetic flux\cite{hawl95}.  Recent numerical experiments\cite{bai13} suggest that a poloidal magnetic pressure as small as 0.1\% of the gas pressure, coherent over scales comparable to the disk thickness, is enough to enhance angular momentum transport by a substantial factor.  Small patches of magnetic flux could arise through statistical fluctuations as a result of local MRI\cite{beck11} but larger coherent fields probably need to be accumulated by advection of flux from larger distances.  Whether efficient flux advection is possible is highly uncertain\cite{lubo94}, and probably depends sensitively on details of the distant outer boundary conditions\cite{bege14} and vertical disk structure\cite{guil12}. Sufficiently coherent poloidal magnetic flux, threading the innermost regions of the rotating accretion flow, is an essential element for extracting rotational energy from the black hole, and presumably also for producing a strong disk wind.   

\item \underbar{\it Black hole spin}.  Part of the gravitating mass $M$ of a spinning black hole, as perceived by a distant observer, is contributed by the spin energy.  In principle, all of this energy --- up to $0.29M$ for an extreme Kerr hole --- can be extracted.  In practice this can be done using coherent poloidal magnetic fields, through the Blandford--Znajek (BZ) process\cite{blan77}.  Since the power extracted is proportional to the square of the net magnetic flux threading the hole as well as the square of the spin parameter $a$, the efficiency of the BZ effect is sensitive to the amount of flux that can be accumulated and held in place against the black hole.  Analytic calculations and relativistic magnetohydrodynamic simulations show that the efficiency can be appreciable, possibly even exceeding the efficiency of the accretion process\cite{reyn06,tche10}.

\end{itemlist}

\section{Modes of Accretion}

The dynamical properties of black-hole accretion flows are affected most strongly by their angular momentum distributions and secondarily by the efficiency of radiative losses.  In almost all astrophysically interesting circumstances the flow is expected to be {\it centrifugally choked}, meaning that the specific angular momentum of the gas supply exceeds a small multiple of $GM/c$.  Various possible modes of accretion, discussed on more detail below, are summarized in Fig.~1.

\begin{figure}[t]
\begin{center}
\psfig{file=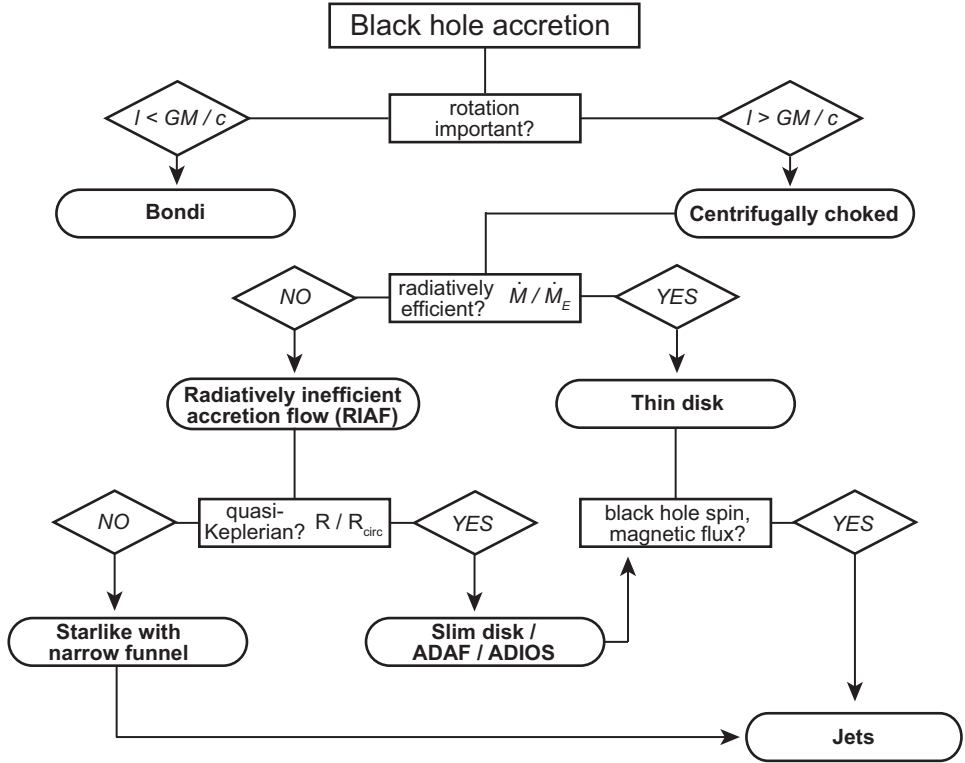,width=5in}
\end{center}
\caption{Flow chart illustrating the main factors determining the dynamical state of black hole accretion flows.  The specific angular momentum relative to $GM/c$ determines whether the flow is centrifugally choked or not, and the same quantity, relative to the Keplerian angular momentum, determines whether the flow resembles a disk or a quasi-spherical (``starlike") envelope.  Radiative inefficiency is a necessary but not sufficient condition for strongly sub-Keplerian rotation.  Additional factors, such as net magnetic flux and the spin of the central black hole, mainly affect the rate of angular momentum transport and the production of jets and winds.}
\label{aba:fig1}
\end{figure}

Hypothetical flows which are not centrifugally choked should resemble spherically symmetric Bondi accretion\cite{bond52}, which may or may not have a high radiative efficiency, depending on the presence or absence of a dynamically significant magnetic field.   

\subsection{Disklike Accretion}

Without significant pressure support, gas will follow approximately Keplerian orbits.  Since energy is generally thought to be dissipated more quickly than angular momentum is transported away, a radiatively efficient flow will circularize to form a thin disk\cite{shak73}. Gas then slowly spirals towards the black hole as it gives up angular momentum, until it reaches the ISCO and plunges into the hole with relatively little additional dissipation.  The energy dissipated at every radius is radiated away locally, and is a well-defined function of mass accretion rate $\dot M$, radius and black hole spin that can be converted into a run of effective temperatures and (assuming a large disk optical depth, which usually applies) a continuum spectrum.  A key strength of accretion disk theory is that these results do not depend on the angular momentum transport mechanism.  Fitting the continuum spectrum close to the inner edge of the disk is the basis of the continuum fitting method for determining black hole spins\cite{mccl14}.  However, portions of the spectrum originating farther out often seem hotter than predicted, probably due to irradiation of the disk, poorly understood optical depth effects, or nonlocal dissipation.

At lower radiative efficiency, the disk will thicken as more internal energy is retained, leading to a {\it slim disk}\cite{abra88}, in which radial pressure forces depress the angular momentum at every radius below the Keplerian value and radial advection of energy becomes appreciable.  

It is important to note that the energy dissipated locally in an accretion disk is {\it not} the same as the gravitational binding energy liberated locally.  Far from the black hole where the angular momentum is close to the Keplerian value, the dissipation rate is three times the local rate of energy liberation because two-thirds of the dissipated energy is transported from closer in by the same torques that transport angular momentum outward.  Overall energy conservation is maintained because the dissipation rate close to the ISCO is lower than the local rate at which energy is liberated.  This means that the outer parts of an accretion disk will accumulate internal energy and becomes unbound unless at least two-thirds of the dissipated energy is radiated away.    

This problem was noted by Narayan \& Yi\cite{nara94,nara95} in their models for {\it advection-dominated accretion flows} (ADAFs), which assume steady accretion with low radiative efficiency.  Self-similar ADAF models exhibit positive Bernoulli function $B$ (where $B$ is the sum of gravitational potential energy, kinetic energy and specific gas enthalpy), implying that elements of gas in the flow are able to unbind neighboring elements by doing work on them.  Eventually the entire flow should disperse.  Possible resolutions that preserve disklike structure include {\it adiabatic inflow-outflow solutions} (ADIOS)\cite{blan99}, {\it convection-dominated accretion flows} (CDAFs)\cite{nara00,quat00}, and models with large-scale matter circulation\cite{bege12}.  These models operate by suppressing the accretion rate relative to the mass supply and/or providing an escape route for energy and angular momentum that avoids the accreting gas.  All of them require some unspecified mechanism for separating accreting matter from outflowing mass, energy and angular momentum.  Despite these uncertainties, various numerical simulations\cite{ston99,ston01,hawl02} suggest that such a separation can take place naturally.

\subsection{Starlike Accretion}

The dynamical significance of the Bernoulli parameter can be demonstrated by considering a sequence of two-dimensional, axisymmetric, self-similar models of quasi-Keplerian rotating flows with pressure. Quasi-Keplerian here means that the specific angular momentum scales with radius as $R^{1/2}$.  If $B$ is either constant or has a radial scaling $\propto 1/R$, then as $B$ approaches zero from below the surface of the disklike flow closes up to the rotation axis, and the flow becomes {\it starlike}.\cite{nara94,blan04} If $B$ becomes positive, the flow becomes unbounded.  This suggests that disklike ADAF models without some kind of escape valve (such as outflow or circulation) are untenable.  Could such flows alternatively resolve their energy crisis by expanding and becoming nearly spherical?

It turns out that the ratio of the specific angular momentum to the Keplerian angular momentum is the parameter that decides whether an accretion flow is disklike or starlike.  Any of the disklike, radiatively inefficient flows have relatively ``flat" density distributions as a function of radius, $\rho \propto R^{-n}$ with $1/2 < n < 3/2$.  But in order for such flows to remain in dynamical equilibrium with $B < 0$, the specific angular momentum has to maintain a fairly large fraction of the Keplerian value.  For the interesting case of a radiatively inefficient accretion flow (RIAF) dominated by radiation pressure, meaning that the adiabatic index $\gamma$ is $4/3$, the minimum specific angular momentum compatible with disklike flow ranges between 74\% and 88\% of Keplerian, as $n$ ranges from 3/2 to 1/2.  Any lower value of angular momentum leads to starlike flow. (This result is sensitive to $\gamma$, with the constraint relaxed for larger values of this parameter. For the limiting case of a gas pressure-dominated RIAF with $\gamma = 5/3$ and $n=3/2$, disklike flow is formally possible for any specific angular momentum, but even in this limit an angular momentum 30\% of Keplerian forces the disk surface to close up to within $5^\circ$ of the rotation axis.)
       
The outer boundary conditions for the mass supply probably play the dominant role in determining whether the angular momentum is large enough to keep the flow disklike.  Gas that is supplied from the outside, in the form of a thin disk, probably holds on to enough angular momentum to remain disklike all the way to the black hole.  Accretion flows in X-ray binaries fall into this category.  On the other hand, flows that start out with a finite supply of angular momentum, such as debris from tidal disruption events or the interiors of stellar envelopes, might become starved of angular momentum (particularly if the envelope expands as it absorbs energy released by accretion) and forced into a starlike state.   
 
The presumption for disklike flows is that the accretion rate adjusts itself so that the circulation, outflow, convection --- or whatever it is that relieves the energy crisis --- is able to carry away any accretion energy that isn't lost to radiation.  This is possible because the radial density and pressure gradients are rather flat, giving the outer flow a high ``carrying capacity" for excess energy, compared to the inner flow where most of this energy is liberated.  But this presumption fails in the case of starlike accretion flows, where the density and pressure profiles are forced to steepen in order to keep the gas bound while satisfying dynamical constraints.  This means that matter is more centrally concentrated around the black hole in a starlike flow, and if this matter is swallowed with an energy efficiency typical for the ISCO it will liberate much more energy than can be carried away by the outer parts of the flow. 

One can envisage several outcomes of this situation: 1) there is no equilibrium configuration, and the flow either blows itself up or becomes violently unsteady, with successive episodes of energy accumulation and release; 2) the gas close to the black hole is pushed inward by the pressure until it approaches the marginally bound orbit, in which case the large accretion rate releases very little energy; and 3) the inner flow finds a way to release most of the accretion energy locally, without forcing it to propagate through the outer flow.  This third possibility, for example, could involve the production of powerful jets that escape through the rotational funnel.  It is hard to decide by pure thought which, if any, of these possibilities is most likely, and simulations have not yet addressed this problem.  Nevertheless, there are observational indications that some tidal disruption events, at least, choose the third option (Sec.~4.3).                

\subsection{Causes of Radiative Inefficiency}

The dynamical properties of black hole accretion flows depend on whether the gas radiates efficiently or not, but do not depend on the mechanisms that determine radiative efficiency.  From an observational point of view, however, these details are crucial because radiatively inefficient flows can be either very faint or, paradoxically, very luminous.  The fiducial mass flux that governs radiative efficiency is related to the Eddington limit, which is the luminosity (assumed to be isotropic) at which radiation pressure force balances gravity for a gas with opacity $\kappa$: $L_{\rm E} = 4\pi GMc/\kappa$.  The accretion rate capable of producing such a luminosity is $\dot M_{\rm E} = L_{\rm E}/ \varepsilon c^2$, where $\varepsilon$ is the radiative efficiency of accretion.  Radiation escaping from such a flow will exert outward pressure forces competitive with centrifugal force, thus creating the dynamical conditions prevalent in radiatively inefficient flow.  In fact, flows with $\dot M > \dot M_{\rm E}$  are literally radiatively inefficient because radiation is trapped and advected inward by the large optical depth at radii $R < (\dot M / \dot M_{\rm E}) GM/c^2$ (to within a factor $\sim \varepsilon$).\cite{bege79} 

Thus, accretion flows with high accretion rates are radiatively inefficient because they liberate more energy than they can radiate, but they are also very luminous because they radiate at close to the Eddington limit.  Such flows are strongly dominated by radiation pressure and should be modeled using an adiabatic index of 4/3.

Accretion flows with low accretion rates can also be radiatively inefficient, because their densities are so low that radiative processes (which typically scale as $\rho^2$) cannot keep up with dissipation (which scales as $\rho$).  Because electrons cool much more rapidly than ions, such flows are expected to develop a ``two-temperature" thermal structure, with $T_{\rm i} \gg T_{\rm e}$.\cite{rees82}  They would be optically thin, dominated by thermal gas pressure, and characterized by an adiabatic index 5/3.  If thermal coupling between ions and electrons is provided by Coulomb interactions, such flows can exist for accretion rates $\dot M < \alpha^2 \dot M_{\rm E}$ (very roughly), where $\alpha\ll 1$ is a widely-used parameter that describes the rate of angular momentum transport.\cite{shak73}

At high ($> \dot M_{\rm E}$) and intermediate ($\alpha^2 \dot M_{\rm E} < \dot M < \dot M_{\rm E}$) accretion rates, the thermal state of accretion is uniquely determined by $\dot M$.  But flows with low accretion rates may exist in either a radiatively efficient (thin disk) or inefficient (two-temperature) state.  It is not understand what would trigger a thin disk in this regime to transition to the radiatively inefficient state, or vice-versa, but there is evidence that such transitions do occur in X-ray binaries. 

\section{Phenomenology of Black Hole Accretion}

\subsection{X-Ray Binaries}

X-ray binaries (XRBs) are the observational manifestations of stellar-mass black holes undergoing disklike accretion of matter captured from a companion star.  The mean rate of mass supply is typically far below $\dot M_{\rm E}$, but the outer disks undergo occasional instabilities\cite{meye81} that dump matter into the central regions at a rate that can approach or even exceed the Eddington limit.  These outbursts can last months, and usually follow a specific sequence of thermal and spectral ``states" that has never been adequately explained.  A schematic representation of the sequence of states in the luminosity--spectral ``hardness" plane is shown in Fig.~2\cite{fend04,bege14}, which is annotated to illustrate a possible model for the origin of the observed hysteresis.

\begin{figure}[t]
\begin{center}
\psfig{file=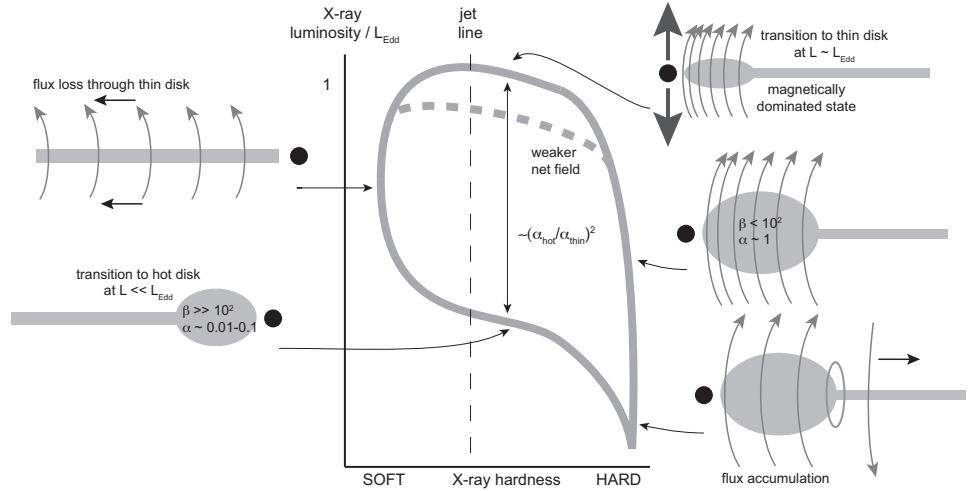,width=5in}
\end{center}
\caption{X-ray binaries during outburst exhibit a temporal sequence of spectral states in the plane of X-ray hardness vs.~luminosity.  A similar sequence is seen in the plane of rms variability vs.~luminosity, with variability and hardness strongly correlated.  The sense of the cycle is counterclockwise, implying that a hard spectrum with increasing luminosity precedes a rapid transition to a soft spectrum at roughly constant luminosity, etc.  The origin of the hysteresis exhibited by the cycle, in which the hard to soft transition occurs at a higher luminosity than soft to hard, is an unsolved problem.  Cartoons illustrating different stages of the cycle show the geometric/thermal characteristics of each state (thin radiatively efficient disk vs. radiatively inefficient hot flow), which are widely agreed on, along with a more speculative proposal for the correlated behavior of the magnetic flux.\cite{bege14} Figure is reproduced from Ref.~12.}
\label{aba:fig1}
\end{figure}

Although the cyclic evolution of XRB outbursts is not well understood, there is reasonable confidence about the physical nature of each state.\cite{remi06,done07} As the luminosity increases at the start of the outburst, the observed X-rays have a very ``hard" (i.e., biased towards high energies), ``nonthermal" spectrum that looks nothing like that expected from a thin (radiatively efficient), optically thick accretion disk.  This is usually interpreted as evidence for a radiatively inefficient, two-temperature flow occupying the region around the black hole. As the accretion rate increases, this hot region shrinks and collapses to form a radiatively efficient disk.  Both the pre- and post-collapse states are highly variable, and show evidence for a relativistic jet that is thought to indicate strong magnetic activity.  Whether this activity is tied to the accumulation of magnetic flux is a subject of current speculation.\cite{dext14,bege14}  A hard spectral ``tail," combined with evidence of thermal disk emission, suggests that a highly magnetized corona surmounts the disk.  Only at the highest luminosities, exceeding a few percent of the Eddington limit, does the magnetic activity calm down (often following a dramatic, final outburst) and the flow resemble a standard thin disk with a thermal spectrum peaking at moderate X-ray energies. 

As the accretion rate declines, the flow does not retrace its previous evolution.  Rather than developing signs of strong magnetic activity and evaporating to form a hot,two-temperature state, the disk remains thin and rather quiescent, decreasing in luminosity while remaining radiatively efficient.  The transition back to a hot accretion state with strong magnetic activity does not occur until the luminosity has dropped to a fraction of the luminosity at which the forward transition took place. This hysteresis, illustrated in Fig.~2, also occurs in XRBs containing a neutron star rather than a black hole,\cite{muno14} so it is apparently intrinsic to the accretion flow and not the properties of the central object. 

\subsection{Active Galactic Nuclei}

An active galactic nucleus (AGN) is produced when a central supermassive ($M> 10^6 M_\odot$) black holes accretes matter from its host galaxy.  The outer boundary conditions for AGN accretion flows are much less understood (and probably much more diverse) than those for XRBs.  Any standard thin disk model set up to supply the requisite mass for a luminous AGN is predicted to become self-gravitating beyond a fraction of a parsec, suggesting that star formation would interrupt or at least modify the flow.\cite{shlo89,good03}  Attempts to resolve this problem --- such as thermal regulation of the disk through stellar processes, magnetically supported disks,\cite{bege07,gabu12} and stochastic injection of matter in random orbits\cite{king07} --- show varying degrees of promise, but none has been strongly supported by observations.  

The diversity of boundary conditions, as well as the cooler temperatures that typically prevail in AGN accretion flows, presumably lead to the highly diverse observational manifestations of these objects.\cite{krol99} Many AGN are apparently surrounded by a geometrically thick torus of opaque, dusty gas that obscures a direct view of the inner accretion flow and reprocesses its emission to longer wavelengths.  Unobscured AGN often exhibit an apparently thermal emission component peaking in the ultraviolet or soft X-rays, as predicted for thin disks with luminosities of roughly a few percent Eddington around black holes in this mass range.  Such AGN also exhibit Doppler broadened emission lines that result from reprocessing of the thermal UV and X-rays, as well as corresponding absorption lines whose large blueshifts indicate that they are produced in a cool, fast wind. 

Other AGN appear to be rather radiatively inefficient, with little evidence for thermal disk emission and luminosities well below the Eddington limit.  Such AGN are primarily detected through broadband emission from powerful, relativistic jets.\cite{bege84}  These are especially well-studied in the radio (mainly because radio telescopes are especially sensitive compared to telescopes operating in other bands), but also emit strongly in all other bands ranging up to X-rays and gamma-rays.  The relativistic nature of these jets is established through clear signatures of Doppler beaming (e.g., one-sided structure on the sky and rapid variability) and apparent superluminal motion (an illusion due to light travel-time effects) in jets pointing close to the line of sight.    

AGN exhibit variability but, unlike XRBs, seldom change their accretion state dramatically.  It could be that the manner in which gas is supplied determines the long-term mode of accretion, e.g., disklike and radiatively efficient when interstellar clouds accumulate and settle into a disk, hot (two-temperature) and perhaps starlike when matter is supplied from a hot halo of gas surrounding the nucleus of the galaxy or from stellar winds, as is believed to be the case in the Galactic Center.\cite{cuad06}  The fact that radiatively inefficient flows with powerful jets occur much more frequently in elliptical galaxies (which have predominantly hot interstellar matter) than in spirals (which have an abundant cool gas) seems to support this environmental interpretation.  These differences in accretion mode could also affect the accumulation of magnetic flux, thus governing jet production indirectly.\cite{siko13}  But one cannot rule out the possibility that AGN undergo state transitions analogous to XRBs, as their vastly larger spatial scales suggest that such transitions would take place too slowly to detect.

\subsection{Tidal Disruption Events}

Tidal disruption events (TDEs) are transient episodes of accretion triggered when a star is partially or completely disrupted by the tidal gravitational field of a supermassive black hole.\cite{hill75,lacy82}  In a typical event about half the mass escapes, but the other half falls back gradually over time, with a rate scaling roughly as $t^{-5/3}$ at late times.\cite{rees88}  

Only black holes less massive than about $10^8 M_\odot$ are capable of disrupting main-sequence stars; more massive holes would swallow such stars whole.  A solar-type star would have to venture to within about 30 Schwarzschild radii of the Galactic Center's black hole (which has a mass of several million $M_\odot$) in order to be disrupted, implying that the mean specific angular momentum of the debris would only be a few times $GM/c$.  This means that fallback cannot lead initially to a disklike accretion flow that extends to large radii, because there is not enough angular momentum compared to the Keplerian value.  At later times, however, as matter is swallowed leaving behind most of its initial angular momentum, the mean angular momentum per unit mass of the remaining material increases and the flow can evolve toward a more disklike state. 

Despite the relatively small cross-section for tidal disruption, TDEs are expected to be fairly common in nearby galaxies, occurring once every $10^4$ years or so in a galaxy like the Milky Way.  About two dozen candidate TDEs have been identified through the soft X-ray thermal spectra predicted to characterize their accretion disks and their characteristic light curves, which peak days to months after the disruption (depending mainly on the mass of the black hole) and then decline roughly according to $t^{-5/3}$.\cite{komo99, geza12}

Disruptions of solar-type stars by black holes in the mass range $10^5 - 10^6 M_\odot$ are predicted to lead to fallback rates large enough to produce super-Eddington luminosities for $\sim 1-3$ yr, if the debris were accreted in real time.\cite{evan89,guil13}  Two TDE candidates have been discovered which appear to exceed the Eddington limit by about two orders of magnitude (for the estimated SMBH mass of $\sim 10^6 M_\odot$), even after correcting for beaming.\cite{burr11,bloo11,leva11,cenk12}  Their observed decay rates suggest that the luminosity tracks the fallback rate, and the presence of a radio afterglow in both cases suggests the production of a relativistic jet. 

If these events represented disklike accretion, one would expect self-regulation of the mass flux reaching the black hole to a value that did not exceed $\dot M_{\rm E}$ by a large factor.\cite{loeb97,stru11} But super-Eddington TDE accretion flows are probably starlike, given the low specific angular momentum of the accreting matter and the fact that it is probably pushed out to rather large distances by the pressure of trapped radiation.  In Sec.~3.2 we argued that radiatively inefficient, starlike flows are unable to regulate the rate at which matter reaches the black hole, but were unable to decide the outcome of the energy crisis that likely ensues.  Observations of super-Eddington X-ray luminosities and jets from TDEs suggest that, at least in these systems, the excess energy finds a relatively stable escape route through the rotational axis.\cite{coug14}  
 
\subsection{Gamma-Ray Bursts}

The close coincidence between long-duration gamma-ray bursts (GRBs) and core collapse supernovae supports the collapsar model, in which the burst results from the formation and rapid growth of a black hole or neutron star at the center of a massive stellar envelope.\cite{woos93} The long duration of such bursts (minutes or more) implies sustained accretion at an extremely high rate,\cite{pira05} while the large total energies involved favor a black hole engine, at least for the most luminous bursts.  The inferred accretion rates initially can be as large as a tenth of a solar mass {\it per second}.\cite{lind10}.  While enormously super-Eddington (by up to 14 orders of magnitude), the initial episode of accretion is not necessarily radiatively inefficient, because neutrinos can carry away most of the liberated energy; however, as the accretion rate declines and neutrino losses become insignificant, the flow must revert to an extremely radiatively inefficient state.

Our ignorance about the angular momentum distribution inside the stellar progenitor make it difficult to determine whether the late-time accretion flow during a long-duration burst is disklike or starlike.  The weight and optical depth of the overlying envelope certainly prevent the Eddington limit from being a factor in regulating the rate at which mass reaches the black hole --- radiation is too thoroughly trapped.  However, we would expect a disklike flow to adjust so that some outward advection or circulation of energy suppresses the accretion by about a factor $(c_s/c)^2$ below the Bondi value, where the latter is calculated using the self-consistent value of the density and sound speed $c_s$ at the ``Bondi radius" $GM/c_s^2$ inside the stellar envelope.  But the accretion rates needed to explain the prompt emission from long-duration bursts far exceed this, suggesting that no such regulation occurs.  A starlike accretion flow, producing orders of magnitude more energy than can be wicked away by the accretion flow, would suffer a similar energy crisis to that in the jetted TDEs.\cite{coug14}.  Like the TDEs, GRBs are evidently able to dispose of the excess energy through powerful jets punching through a quasi-spherical envelope.

These jets are remarkable for their enormous bulk Lorentz factors ($\Gamma\sim 100-1000$) which are inferred from variability considerations and the requirement that the gamma-rays be able to escape.  These Lorentz factors are 1--2 orders of magnitude higher than those found in other jets produced by black hole accretion, such as the jets from AGN and XRBs.  I will comment on the possible significance of this below.

\section{Unsolved Problems}

Here is a selection of problems that are currently of interest to workers in the field of black hole accretion.  These are problems that particularly interest me, and I make no claim that they are widely agreed upon as among the most pressing issues. 
 
\begin{itemlist}
\item \underbar{\it What triggers state transitions}?  

While there is fair agreement about the nature of each XRB state, the factors that cause transitions between states are much less clear.  In particular, the transition from the thin disk state to the hot, radiatively inefficient (and presumably two-temperature) state, is poorly understood.  Both states are possible for a range of accretion rates spanning many decades, and there is a large ``potential barrier" against converting a thin disk to a hot torus, because the high density of the thin disk keeps electrons and ions extremely tightly coupled.  Some kind of bootstrap evaporation process may be necessary, as has been proposed for cataclysmic variables.\cite{meye94} Formation of coronae above thin disks has been observed in magnetohydrodynamic simulations with radiative cooling\cite{jian14}, but the effects of microphysical plasma processes (such as viscosity and electrical resistivity\cite{balb08}), which may regulate the level of magnetic activity, have yet to be assessed.

The question of what triggers state transitions is tied up with the question of what causes the hysteresis (Fig.~2), in which the transition from a hard (radiatively inefficient) to a soft (radiatively efficient) state occurs at a higher luminosity than the reverse transition.  Apparently there is a second parameter (in addition to $\dot M/\dot M_{\rm E}$) affecting the transition, and this parameter is correlated with location in the cycle.  The magnetic flux threading the disk, which can depend on accretion history, is an attractive candidate,\cite{petr08,bege14} as is the magnetic Prandtl number, the ratio of viscosity to resistivity.  \cite{pott14} Other, less orthodox mechanisms, such as the history of disk warping, may also play a role.\cite{nixo14} 

\item \underbar{\it What stabilizes radiation-dominated accretion disks}? 

The simple $\alpha$-parameterization of angular momentum transport in thin accretion disks\cite{shak73} predicts that disks dominated by radiation pressure should be thermally and viscously unstable,\cite{shak76} i.e., they should heat up and thicken while clumping into rings. Yet the luminous, thin-disk states of XRBs show surprisingly little variability,\cite{gier04} suggesting that this prediction is not borne out.  This would not be surprising, given the oversimplification inherent in the $\alpha$-model, were it not the case that simulations appear to show at least the kind of thermal instability predicted\cite{jian13} (the models have not been run long enough to check viscous instability).  Models to address this problem rely on providing an extra channel for release of energy, e.g., through disk turbulence\cite{zhu13} or winds\cite{libe14}, and/or diluting radiative pressure support for the disk, e.g., through magnetic fields.\cite{bege07} Until we understand the stability properties of luminous disks, it will be hard to develop a compelling explanation for state transitions.

\item \underbar{\it How limiting is the Eddington limit}?  
 
The Eddington limit is often regarded as an upper limit to the luminosity of an accreting black hole, and therefore as setting an upper limit to the mass accretion rate.  In disklike flows with a super-Eddington mass supply, such as the XRB SS 433, there are indeed reasons to believe that most of the mass flux in excess of $\dot M_{\rm E}$ is reversed and flung away before getting anywhere near the black hole.  The most luminous quasars also seem to respect the Eddington limit, although it is possible that this could result from a selection effect (super-Eddington quasars might be sufficiently obscured to have escaped identification) or a coincidence due to a distribution in mass supply rates that decreases rapidly with $\dot M$.  But we have seen that in starlike flows, such as the super-Eddington phases of TDEs, such self-regulation may be impossible. And GRBs clearly violate the Eddington limit by many orders of magnitude.  
  
Exceeding the Eddington accretion rate should be distinguished from violating Eddington's limiting luminosity. Black holes can grow at a rate that exceeds $L_{\rm E}/c^2$ for a number of reasons.  For one, radiation produced in an accretion flow need not escape; it can be trapped and swept through the horizon.\cite{bege79}  This is one aspect of accretion that truly distinguishes black holes from other compact objects such as neutron stars.  For another, the matter responsible for liberating most of the accretion energy is not necessarily the same matter that radiates it away.  In a flow with large density contrasts on small scales, radiation could ``go around" opaque clumps, escaping mainly through the low-density interstices.  But if the low- and high-density regions are tied together by magnetic tension, a super-Eddington escaping flux will not exert enough force to stop accretion.\cite{shav98, bege06} For a third, some gas in a radiatively inefficient accretion flow could be swallowed with low binding energy, allowing a large amount of accreted matter to release relatively little radiation.  

From the point of view of escaping luminosity, the Eddington limit is a spherical idealization, sensitive to geometric modifications.  To give a simple example, a geometrically thickened disk, its surface a cone about the rotation axis, can exceed the Eddington luminosity by a factor proportional to the logarithm of the  ratio between the outer and inner radii.\cite{king08}  The apparent luminosity is enhanced by an additional factor because this luminosity is focused into  the solid angle subtended by the disk surface.  Inhomogeneities in the disk density, such as those caused by ``photon bubbles," can also create intrinsically super-Eddington luminosities.\cite{bege06}

More subtle are global effects that make accretion flows unable to regulate their power outputs, such as the steep density profiles expected to develop in starlike accretion flows.  Here we have no obvious theoretical reason to demand that the flow rid itself of the excess energy in an orderly way; it would be quite plausible if the flow blew itself apart.  But the evidence from super-Eddington TDEs, and possibly from GRBs, indicates that somehow a pair of jets is able to carry off the excess energy.  How efficiently this kinetic energy is converted into radiation is another challenging unsolved problem. 
   
\item \underbar{\it What causes quasi-periodic oscillations}?   

Power density spectra of XRB variability display an array of rather narrow peaks, called {\it quasi-periodic oscillations} (QPOs), which often contain a significant amount of power (a percent or more) in all but the thermal, thin disk states.  High-frequency QPOs typically have frequencies of hundreds of Hz, and are presumably associated with dynamical processes (orbital motions, p-mode oscillations, etc.) in the inner portions of the accretion flow.  A 3:2 resonance which is often seen can be interpreted in terms of epicyclic motions in a relativistic gravitational field.\cite{abra01} 

Low-frequency QPOs, with frequencies of less than 0.1 to a few Hz, are more mysterious because they are unlikely to be associated with disk dynamics very close to the black hole, where most of the energy is liberated, yet they can carry a significant fraction (up to a few percent) of the total accretion luminosity.  Both kinds of QPOs are primarily a feature of the hard X-ray spectral bands, which is believed to be produced by a hot, radiatively inefficient (two-temperature) accretion flow in low luminosity states, and  a corona surmounting a thin, magnetically active disk in high-luminosity states.  For the low-frequency QPOs, this suggests that energy is being spread through the inner hot region by nonradiative processes, to be modulated by some kind of oscillation or rotational motion at a rather sharp outer boundary. Candidates for the modulation mechanism include coherent Lense-Thirring precession of a hot accretion flow with an outer radius of several tens of gravitational radii\cite{ingr09} and rigid rotation of an extended ($\sim 10^3 R_g$) magnetosphere,\cite{meie12} which requires an enormous amount of trapped magnetic flux.  Other suggestions appeal to thermal or viscous timescales, but no model has been particularly satisfactory to date. 

\item \underbar{\it Are jets always propelled by coherent magnetic fields}?  

According to standard theories for jet formation, the power of a jet depends on the total magnetic flux threading the engine, $L_{\rm J} \sim \Phi^2 \Omega^2/c$, where $\Phi$ is the net magnetic flux and $\Omega$ is the angular velocity of the crank.  While the rms magnetic field {\it strength} in the inner region of a TDE accretion flow can be substantial, most of this is associated with turbulent field resulting from the MRI.  The net poloidal {\it flux} is limited to the magnetic flux contained in the disrupted star, which is too small to power the observed jets by about five orders of magnitude.\cite{tche14} Whether a starlike, super-Eddington accretion flow can generate large-scale fluctuating fields with enough coherence to power the observed jets is an open question.  GRB jets face a similar shortfall, although the likely discrepancy is only one or two orders of magnitude. 
 
This suggests that jets in TDEs and perhaps GRBs are propelled by the energy in chaotic magnetic fields,\cite{hein00} which would quickly decay by turbulent reconnection into radiation pressure.  Radiative acceleration of optically thin gas to relativistic velocities is severely limited by radiation drag effects, which are made worse by relativistic aberration,\cite{phin82} but these effects can easily be ameliorated if the flow entrains a large enough optical depth in ambient matter to shield itself.  A marginally self-shielded jet could be accelerated to a Lorentz factor that is some fractional power ($\sim 1/4$) of the Eddington ratio $L_{\rm J}/L_{\rm E}$, which could explain why GRB jets are so fast.  

It is curious that the most highly relativistic jets known --- those associated with GRBs --- should emerge from the most optically thick, radiation-dominated regions.  Perhaps this reflects the difficulty that large-scale coherent magnetic fields face in converting most of their energy into motion.  Once they reach moderately relativistic speeds, corresponding to rough equipartition between kinetic energy and Poynting flux, the electric field cancels out nearly all of the accelerating force and the energy conversion process stalls. Some form of magnetic dissipation may be necessary to catalyze further conversion of Poynting flux into kinetic energy.\cite{lyub09,tche09}  

These arguments also apply to the jets which are more likely to be propelled by coherent magnetic flux, i.e., those produced in AGN and XRBs.  It is commonly believed that jets cannot be propelled to extremely relativistic speeds by thermal pressure, but this is true only if most of the energy is quickly transferred to electrons, which cool rapidly. At the low densities and optical depths present in AGN jets, there is no reason to exclude a sizable contribution from relativistic ion thermal pressure.  On the other hand, the fact that these jets seem to be limited to fairly low Lorentz factors, in the range of a few to a few tens, might indicate that such dissipative processes are not very effective.  

\end{itemlist}

\section{Looking Forward}

Our understanding of black hole accretion seems to be in pretty good shape. Magnetohydrodynamic simulations have shown that MRI really can drive accretion, and many other observable phenomena besides, and gives rough scaling relations that are closer to the prescient $\alpha$-model of Shakura and Sunyaev\cite{shak73} than we had any right to expect.  On the other hand, exquisite observations of time-dependence and spectral transformations in XRBs, and completely new regimes of black hole accretion in TDEs and GRBs, show that the phenomenology of black hole accretion is even richer than we thought.  

MHD simulations have come into their own as mature laboratory tools.  Shearing-box experiments have been crucial but are probably approaching the end of their useful life.  Global simulations --- which at the very least are essential for understanding the coupling of disks to winds --- have begun to take their place, but do not yet have adequate dynamic range.  This is presumably just a matter of computer speed and available CPU time.  We will also need to incorporate microphysical effects, such as viscosity, resistivity and heat conduction, which appear to be important for regulating the level of MRI-driven turbulence and perhaps its coupling to large-scale dynamos.  On small turbulent scales, the relevant microphysical effects are likely to be collisional, but we will also need codes that can identify current sheets on large scales and compute the collisionless reconnection that is likely to occur.\cite{uzde13}  Finally, the increasing importance of radiation-dominated flows, with luminosities approaching or even greatly exceeding the Eddington limit, means that radiation magnetohydrodynamic codes will be essential.  Given the likely role of radiation pressure in driving some of the fastest jets, these codes will have to do radiative transfer in regimes where simple closure schemes are likely to fail.    
 
Not all of these tools are in place yet, but there is every likelihood that these advances will happen soon.

\section*{Acknowledgments} I thank Prof.~Phil Armitage for numerous conversations, both over beers and above treeline, and for preparing the figures.  My research on black hole accretion and its manifestations is supported in part by National Science Foundation grant AST 1411879, NASA Astrophysics Theory Program grants NNX11AE12G and NNX14AB375, and Department of Energy grant DE-SC008409.

\end{document}